\documentclass[amsmath,amssymb,aps,showpacs]{revtex4}

\usepackage{hyperref}
\usepackage[utf8]{inputenc}
\usepackage{graphicx}

\begin{document}
\title{On the phase-shift parametrization and ANC extraction from elastic-scattering data}

\author{O.\ L.\ Ramírez Suárez\footnote{Present address: 
Grupo de Simulación, Análisis y Modelado (SiAMo)
Universidad ECCI, Bogotá, Colombia}}
\author{J.-M.\ Sparenberg}
\affiliation{Physique Nucléaire et Physique Quantique, C.\ P.\ 229, Université libre de Bruxelles (ULB), B 1050 Brussels, Belgium}

\date{\today}

\begin{abstract}
We develop a method to parametrize elastic-scattering phase-shifts for charged nuclei, based on Padé expansions of a simplified effective-range function. The method is potential independent and the input is reduced to experimental phase shifts and bound-state energies. It allows a simple calculation of resonance properties and of asymptotic normalization constants (ANCs) of subthreshold bound states. We analyze the $1^-$ and $2^+$ phase shifts of the $^{12}$C$+\alpha$ system and extract the ANCs of the corresponding bound states. For the $1^-$ state, a factor-3 improvement with respect to the best value available today is obtained, with a factor-10 improvement in reach. For the $2^+$ state, no improvement is obtained due to relatively larger error bars on the experimental phase shifts.
\end{abstract}


\maketitle

Low-energy nuclear reactions, in particular those relevant to nuclear astrophysics \cite{Bertulani}, are a fascinating application field for quantum scattering theory \cite{Joachain}.
Because of the Coulomb repulsion between nuclei, cross sections are often impossible to measure directly and theoretical extrapolation are indispensable.
The simplest of these reactions, elastic scattering, is theoretically described with a partial-wave decomposition, each partial wave $l$ being fully characterized by a nuclear phase shift $\delta_l$ or scattering matrix $S_l=e^{2i\delta_l}$, both functions of the energy.
Resonances appear as fast increases of $\delta_l$ at real positive energies or as poles of $S_l$ at complex energies.
For some reactions subthreshold bound states at small negative energies also play an essential role.
This is for instance the case for the $1^-$ and $2^+$ bound states lying just below the $^{12}$C$+\alpha$ threshold which strongly affect the $^{12}$C$(\alpha,\gamma)^{16}$O cross section, an important reaction for stellar evolution \cite{buchmann96}.
For these states, energies are generally well-known experimentally but not the ANC of their wave function.
Hence, various methods have been proposed to indirectly extract these ANCs from experimental data, like $\beta$-delayed $\alpha$ emission \cite{azuma94} or $\alpha$-transfer reactions \cite{PRL83}.

A natural way to extract an ANC for a given partial wave is to parametrize the corresponding experimental phase shifts and to extrapolate this parametrization at negative energies, as bound states also correspond to scattering-matrix poles \cite{Joachain}.
The usual tool to do so is the reaction- (R-)matrix method \cite{RPP71}, which describes both resonant and bound states as poles characterized by real energies and widths (Mittag-Leffler expansion).
This motivated the measurement of high-precision $^{12}$C$+\alpha$ phase shifts \cite{PRL88,PRC79} but led to a loose constraint on the $1^-$ ANC \cite{PRL88} and to a questionable constraint on the $2^+$ ANC \cite{sparenberg04}.
The background phase shifts (between resonances) are indeed described in terms of a channel radius and of a high-energy background pole, which adds several parameters with no direct physical meaning to the fit.
Hence, simpler parametrizations are necessary.

A first option is the effective-range function (ERF) $K_l$ of Eq.\ \eqref{ERF} \cite{NPB60}.
Its analyticity properties imply the existence of a Maclaurin expansion, the effective-range expansion \cite{Joachain}, which provides both a parametrization of $\delta_l$
and an access to subthreshold-bound-state ANCs \cite{PRC29,PRC81}.
This expansion is generally limited to low energies, which restricts the analysis of experimental data \cite{JPCS312},
but this could be overcome by the use of Padé approximants \cite{PRC29,PRC39,blokhintsev93,pupasov11}.
A more serious drawback is that the ERF is a weakly sensitive quantity: very close ERFs can lead to very different, sometimes unphysical, phase shifts and ANCs \cite{JPCS312,orlov16}.
This flaw is due to the second ($h$-)term of Eq.\ \eqref{ERF} and can be simply avoided by directly expanding the first term,
which is proportional to the inverse of the modified K-matrix ${\cal K}_l$ \cite{NPA271,humblet90,mukhamedzhanov99}.
Like the R-matrix, ${\cal K}_l$ is usually Mittag-Leffler expanded.
The background description does not require a channel radius but is however complicated \cite{brune96,humblet98}.
For $^{12}$C+$\alpha$ both methods lead to similar ANC constraints \cite{azuma94}.

In the present work, we simplify things even further by directly expanding function $\Delta_l$ of Eq.\ \eqref{Dlinv},
which can be considered as a simplified $K_l$ (no $h$) or ${\cal K}_l^{-1}$ (no $w_l$) function.
Function $\Delta_l$ is also used in the context of the quantum-defect theory for attractive Coulomb potentials \cite{burke}
and is independent of the partial wave.
This is desirable because the Coulomb potential dominates the centrifugal potential at large distances,
where low-energy scattering properties originate.
We also show that a Padé approximant is more efficient than an Mittag-Leffler expansion, with fewer and more physical parameters, hence leading to better constraints on the $^{12}$C$+\alpha$ ANCs from the phase-shift parametrizations.

We consider the elastic scattering of two particles of charges $Z_1$e and $Z_2$e at positive energy $E=\frac{\hbar^2}{2\mu}k^2$ in the center-of-mass frame, with $k$ the wave number and $\mu$ the reduced mass.
The scattering matrix reads \cite{Joachain,Bertulani}
\begin{equation}\label{Smat}
S^\mathrm{tot}_l=e^{2i\sigma_l} e^{2i\delta_l}
\equiv \frac{\Gamma(l+1+i\eta)}{\Gamma(l+1-i\eta)} \times \frac{\cot \delta_l +i}{\cot \delta_l - i},
\end{equation}
while the usual effective-range function reads \cite{NPB60,PRC29}
\begin{equation}\label{ERF}
K_l=\frac{2w_l}{l!^2a_N^{2l+1}}\left[\frac{\pi \cot \delta_l}{e^{2\pi\eta}-1}+h \right],
\end{equation}
with the Coulomb phase shifts $\sigma_l$, $\eta=1/a_N k$, $a_N= 4\pi\epsilon_0 \hbar^2/\mu Z_1Z_2\text{e}^2$ and $w_l=\prod_{j=0}^l\left[1+(j/\eta)^2\right]$.
Function $h$
is aimed at improving the analyticity properties of $K_l^c$,
the analytic continuation of $K_l$ in the complex plane \cite{NPB60}.

Here we use the simpler function for all partial waves
\begin{equation}\label{Dlinv}
\Delta_l=\frac{2 \pi}{a_N} \frac{\cot \delta_l}{e^{2\pi\eta}-1},
\end{equation}
which reduces to the standard $l=0$ ERF for the neutral case, $k \cot \delta_0$, in the limit $\eta \rightarrow 0$.
Equations \eqref{Smat}-\eqref{ERF} imply that $\Delta_l$ allows to express both the nuclear scattering matrix $S_l$ and the standard ERF in a simple way,
relating them through the cotangent function.
It connects to the modified K-matrix through $\Delta_l = l!^2 a_N^{2l}/w_l \mathcal{K}_l$
(see also Refs.\ \cite{PRC88,PRC63,PRC61} where the function $D_l=4/a_N \Delta_l$ is used instead).
This relation implies that the zeros and poles of both functions are exchanged and that $\mathcal{K}$ is stronger energy-dependent than $\Delta_l$ for $l>0$, which makes it less easy to parametrize.

Let us now study the properties of $\Delta_l$ for positive, zero and negative energies.
For $E>0$, Eqs.\ \eqref{Smat} and \eqref{Dlinv} imply that $\Delta_l$ is real, keeps $S_l$ unitary and has simple poles (resp.\ zeros) at $E_{\infty,j}$ (resp.\ $E_{0,j}$) according to
\begin{equation}\label{Dlinvpoleszeros}
\frac{\delta_l(E)}{\pi/2}=
\begin{cases}
\text{even}, & \text{for } E=E_{\infty,j} \quad (j=1,\dots,N_\infty),\\
\text{odd}, & \text{for } E=E_{0,j} \quad (j=1,\dots,N_0).
\end{cases}
\end{equation}
These zeros and poles characterize the general structure (resonances and background) of the phase shift for positive energies.
For $E=0$, if the scattering length $a_l=-1/K_l(0)$ does not vanish then $\Delta_l$ admits a Maclaurin expansion because $K_l$, $w_l$ and $h$ admit one.
Since $w_l(0)=1$ and $h(0)=0$, one then gets the direct link
$ \Delta_l(0) = -l!^2 a_N^{2l}/a_l$
and the phase-shift behavior $\delta_l\propto e^{-2\pi\eta}$ for $E\rightarrow 0^+$.
For $E\rightarrow \infty$ on the other hand, one has $\delta_l \propto k^{-1}$ and hence $\Delta_l \propto E^{-1}$,
assuming the nuclear potential is regular enough and neglecting relativistic effects \cite{Joachain}. 
From there comes
\begin{equation}\label{Dlinvf}
\Delta_l = \frac{\prod_{j=1}^{N_0}\left[1-(E/E_{0,j})\right]}
{\prod_{j=1}^{N_\infty} \left[1-(E/E_{\infty,j})\right]}g,
\end{equation}
where $g$ is a real analytic function of $E$ for $E\ge 0$, behaving as $g\propto E^{1-N_0+N_\infty}$ for $E\to\infty$.

For $E<0$, $K^c_l$ has to be used.
It is real and satisfies
\begin{equation}\label{ERFcBS}
K_l^c(E_B)=\left.\frac{2w_l}{l!^2a_N^{2l+1}}h^c\right|_{\eta=\eta_B},
\end{equation}
for the bound-state energy $E_B=\frac{\hbar^2}{2\mu}k_B^2$ and the imaginary parameter $\eta_B=1/(a_Nk_B)$ \cite{PRA26,PRC29}.
This means that $\Delta_l^c$, the analytic continuation of $\Delta_l$, simply vanishes at $k=k_B$.
Let us recall three facts of single-channel scattering: (i) $k_B$ is a simple pole of the S-matrix. (ii) If $E\to0^+$ from Eqs.\ \eqref{Smat} and \eqref{Dlinv} one gets $S_l\propto1/\Delta_l$, which agrees with the condition $\Delta_l^c(E_B)=0$ especially for a weakly bound state (which can be interpreted as $E\to 0^-$). (iii) Bound states have the same mathematical properties as weakly bound states.
From there, we conjecture that \emph{$E_B$ is a simple zero of $\Delta_l^c$}.
We thus expect $g$ to be well approximated by the Padé approximant
\begin{equation}\label{gPade}
g_\text{Padé}=
\frac{\sum_{j=0}^N p_jE^j}{\sum_{j=0}^{M} q_j E^j}
\prod_{n=1}^{N_B}\left(1-\frac{E}{E_{B,n}}\right),
\end{equation}
with $p_j$ and $q_j$ real numbers, $N\geq 0$ and
\begin{equation}\label{CoM}
M=N+N_0-N_\infty-1+N_B\geq 0.
\end{equation}
Note that
(i) without loss of generality, we can choose $p_0=1$ which implies $q_0\approx -a_l/l!^2a_N^{2l}$ if $g_\text{Padé}\approx g$ around zero energy,
(ii) Eq.\ \eqref{CoM} shows that the set of free parameters increases linearly with the number of bound states.

We can corroborate our hypothesis ($g_\text{Padé}\approx g$) by replacing Eq.\ \eqref{gPade} in Eq.\ \eqref{Dlinvf} and fitting the sets $\{p_j\}$ and $\{q_j\}$ to experimental data
via Eq.\ \eqref{Dlinv} and \eqref{Smat}.
Once these sets are fitted, one can compute resonances characterized by energies $E_r$s and widths $\Gamma$s by finding the poles $E_\text{pole}=E_r-\frac{i}{2}\Gamma$ of the scattering matrix \eqref{Smat}. Similarly, one can also estimate the ANC for a bound state. The formula that links the ERF with the ANC is presented in Ref.\ \cite{PRC29}; here we write it down in terms of $\Delta_l^c$ as
\begin{equation}\label{ANC}
\text{ANC}=\frac{\Gamma(l+1+|\eta_B|)}{|\eta_B|^l}\left|w_l(\eta_B)\frac{d\Delta_l^c}{dk^2}\right|_{k=k_B}^{-1/2}.
\end{equation}
Note that Eq.\ \eqref{ANC} does not depend on $h^c$, its computation can be done analytically by using Eqs.\ \eqref{Dlinvf} and \eqref{gPade}, and all the information from positive energies is in $\Delta_l^c$.
For weakly bound states,
we expect the following linear approximation to be precise between $E=0$ and $E_B$:
\begin{equation}
 \Delta_l(E) \mathop{\approx}_{E\in [E_B, 0]} \frac{l!^2 a_N^{2l}}{-a_l} \left(1-\frac{E}{E_B}\right).
\end{equation}
From there and Eq.\ \eqref{ANC}, we deduce an approximate expression for the ANC in terms of the scattering length,
\begin{equation}
 \mathrm{ANC} \approx \frac{\Gamma(l+1+|\eta_b|)}{l!} \kappa_b^{l+1} \sqrt{\left| \frac{a_l}{w_l(\eta_b)}\right|}.
\label{ANC_a}
\end{equation}

To test our method, we first applied it to the $^{12}$C+$\alpha$  $d$-wave potential of Refs.\ \cite{sparenberg04,PRC81}, which has $N_B=N_0=3$, $N_\infty=2$ and which displays a bound state at the experimental energy $-245$ keV below the $^{12}$C+$\alpha$ threshold,
with an ANC of $138.4\times 10^5$ fm$^{-1/2}$.
This study \cite{ramirez14} shows on the one hand that a Padé approximant of the $\Delta_2$ function is a very efficient way of parameterizing phase shifts on large energy ranges. For instance, with $N=1, M=4$ in Eq.\ \eqref{gPade}, one can fit the phase shifts of this potential up to 2.4 GeV. This confirms that, with Padé approximants, effective-range expansions are not restricted to low energies anymore.
On the other hand, at low energies, this study shows that such Padé approximants can be used to deduce a subthreshold-bound-state ANC from scattering phase shifts. For instance, with an $N=2, M=4$ fit of the [1-10] MeV phase shifts of the $^{12}$C+$\alpha$ potential, one gets an ANC of $138.2\times 10^5$ fm$^{-1/2}$.
Interestingly, Eq.\ \eqref{ANC_a} provides $130.5\times 10^5$ fm$^{-1/2}$, i.e.\ a 6\% error, to be compared with the 11\% error obtained from Eq.\ (19) of Ref.\ \cite{PRC81}.
Of course, these equations can only be used when the scattering length is known, which is the case for a theoretical model but not when only experimental phase shifts and binding energies are known.

\begin{figure}[]
\scalebox{0.5}{\includegraphics{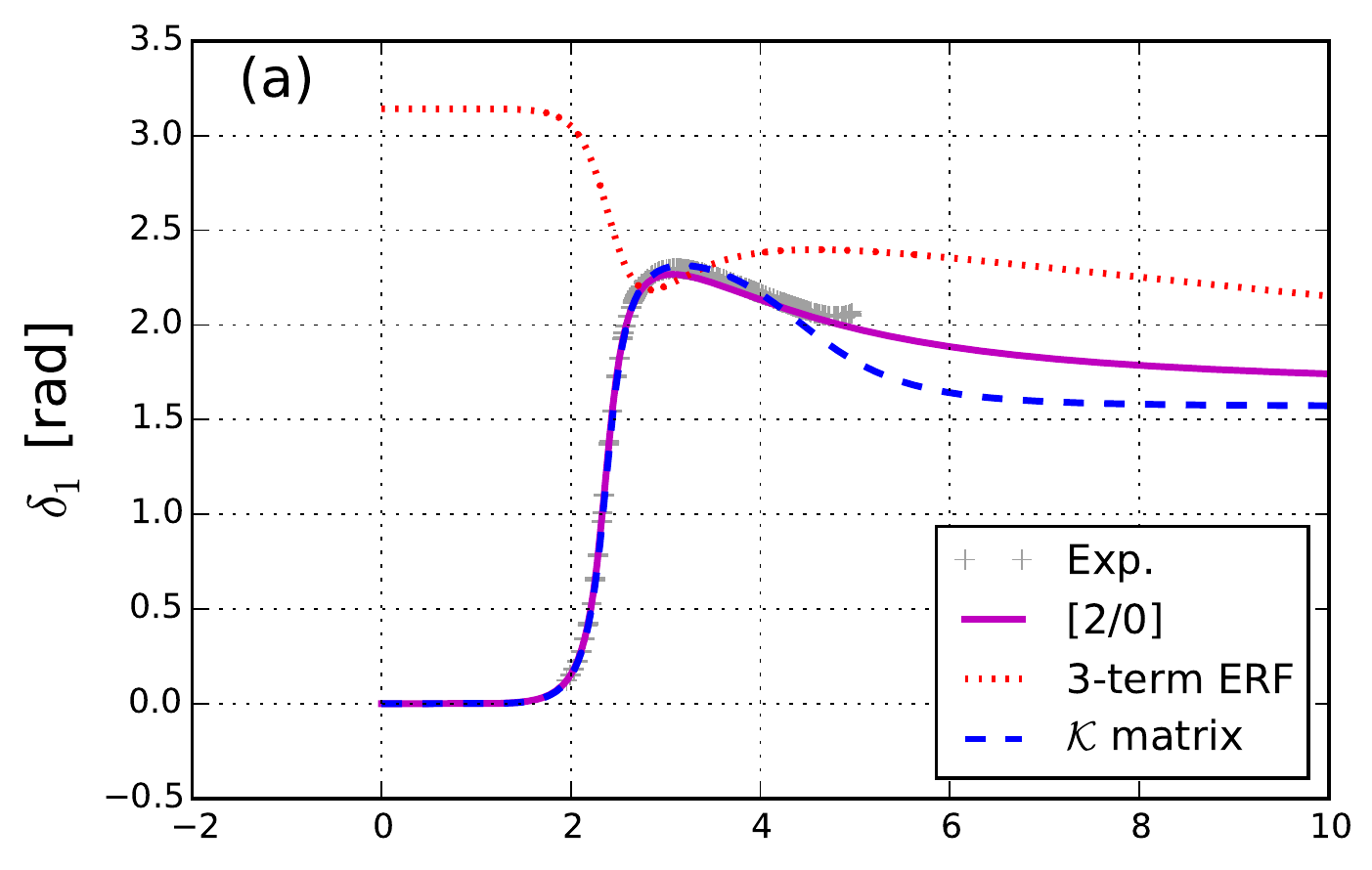}} \\
\scalebox{0.5}{\includegraphics{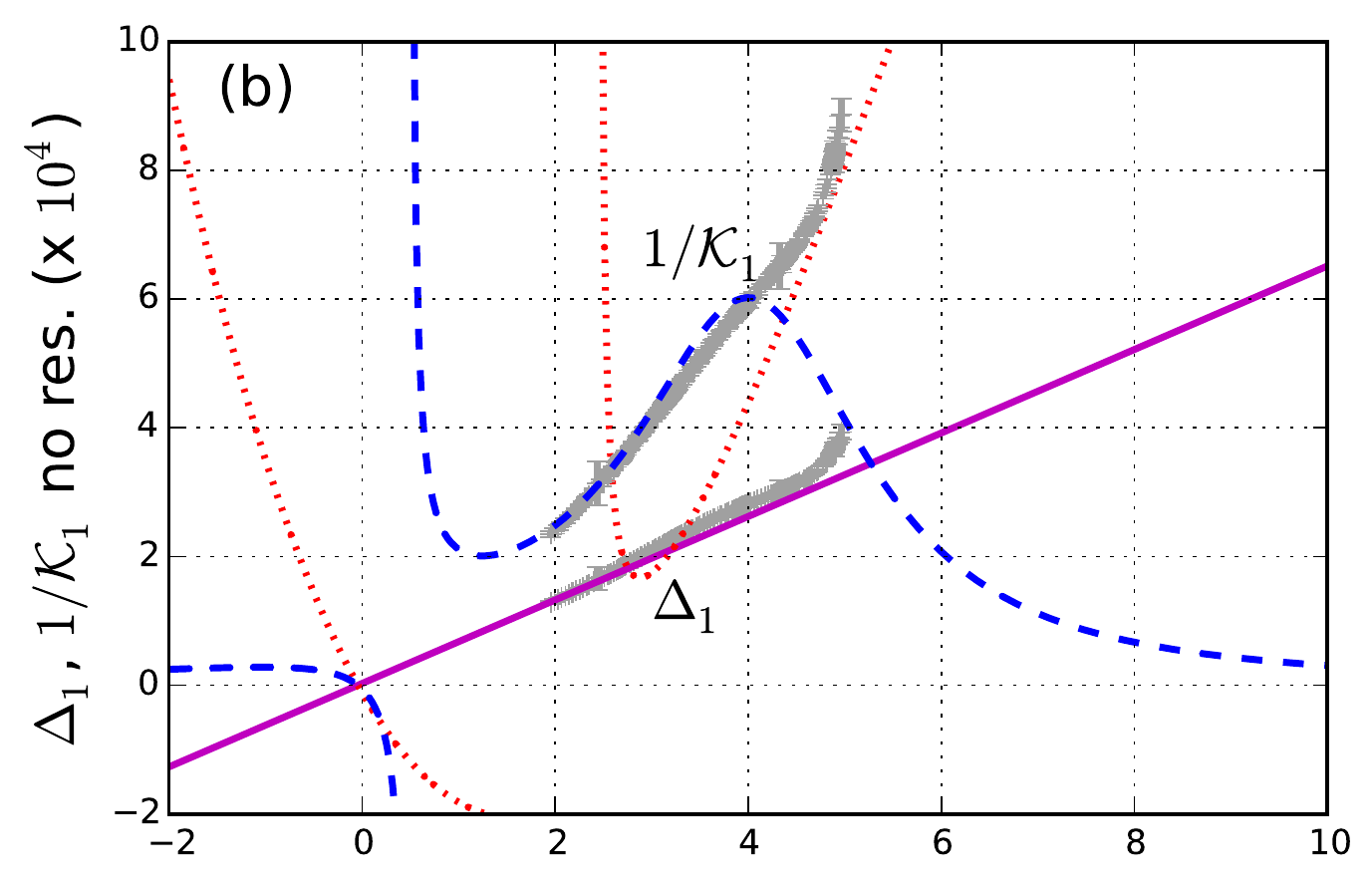}} \\
\scalebox{0.5}{\includegraphics{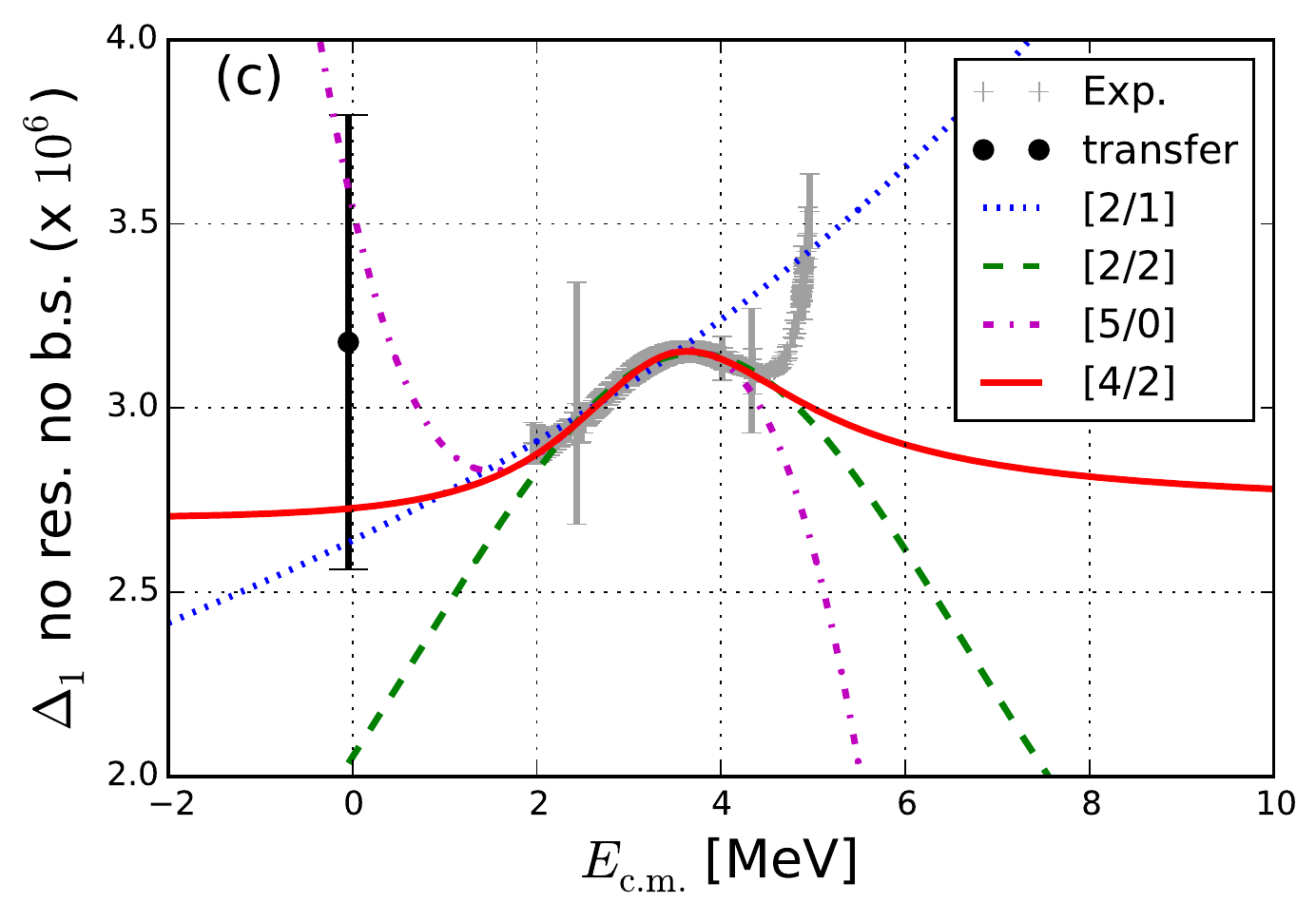}} \\
\caption{$^{12}$C$+\alpha$ elastic-scattering $p$-wave (a) phase-shifts $\delta_1$,
(b) simplified effective-range function (ERF) $\Delta_1$ with resonance zero removed,
(c) $\Delta_1$ with resonance and bound-state zeros removed, compared with the value deduced from the transfer-reaction ANC of Ref.\ \cite{PRL83}.
The experimental data are from Ref.\ \cite{PRL88,PRC79} and the $[N_\mathrm{fit}/M_\mathrm{fit}]$ Padé approximants are detailed in the text.
In (a) and (b), the standard ERF fit of Ref.\ \cite{orlov16} and the modified K-matrix fit of Ref.\ \cite{azuma94} are also shown.} 
\label{fits_pwave}  
\end{figure}

Let us now apply our method to the phase shifts of Refs.\ \cite{PRL88,PRC79}, which have been obtained through an R-matrix fit of high-precision cross sections measured on the energy interval [1.955-4.965] MeV.
For the $1^-$ phase shifts [see Fig.\ \ref{fits_pwave}(a)], the resonance at 2.4 MeV is clearly seen and a threshold effect or the tail of a higher-energy resonance can be guessed above 4.1 MeV.
To simplify the following discussion and since we are interested in low-energy extrapolation towards the subthreshold bound-state energy at $E_B = -0.045$ MeV, we do not take data above 4.1 MeV into account (we have checked that our conclusions are essentially unaffected by keeping them).
The corresponding $\Delta_1$ function behaves like $-6.6929 \times 10^{-5} (E-E_\mathrm{res})$ fm$^{-1}$ for $E\approx E_\mathrm{res} = 2.442$ MeV.
By dividing it by $1-E/E_\mathrm{res}$,
one gets the 'no res.' data of Fig.\ \ref{fits_pwave}(b), which is approximately linear below 4.5 MeV.

Remarkably, extrapolating this data towards negative energy leads to a zero close to $E_B$,
as shown by our [2/0] fit (detailed below).
This suggests that $\Delta_1$ relates the subthreshold bound state to the experimental phase shifts in a very simple way.
This is not the case for ${\cal K}_1^{-1}$, also represented on Fig.\ \ref{fits_pwave}(b): it depends more strongly on the energy and a linear fit does not provide the correct binding energy.
The K-matrix Mittag-Leffler fit of Ref.\ \cite{azuma94} (adjusted on older data) is also represented in Figs.\ \ref{fits_pwave}(a) and (b); though the phase shifts are satisfactory, $1/{\cal K}_1$ presents a complicated structure, due to the background description and to an imposed real reduced width for the bound state.
Our Padé fit is simpler and corresponds to an imaginary reduced width.
Also represented are the $\delta_1$ and non-resonant $\Delta_1$ corresponding to the 3-term effective-range expansion of Ref.\ \cite{orlov16};
they are clearly unphysical.
Interestingly, a plot of $K_1$ leads to practically indistinguishable curves for the three fits presented here, illustrating the lack of sensitivity of the usual effective-range function to low-energy physical quantities.

Since the slope of $\Delta_1$ at the bound-state energy directly provides the ANC, according to Eq.\ \eqref{ANC},
and since the bound-state energy is precisely known experimentally, we again divide $\Delta_1$ by $1-E/E_B$,
which leads to the 'no res.\ no b.s.' data of Fig.\ \ref{fits_pwave}(c).
This reveals even more details.
Ideally, the resulting curve should be constant, which would directly lead to the bound-state ANC.
However, the situation is not that simple:
except for a rather fast variation above 4.1 MeV (much more clearly seen on $\Delta_1$ than on $\delta_1$),
the background $\Delta_1$ function is only {\em approximately} constant:
it varies between 1.15 and $1.28 \times 10^{-6}$,
which corresponds to an ANC of $214(6) \times 10^{12}$ fm$^{-1/2}$.
This is in perfect agreement and already thrice more accurate than the value of Ref.\ \cite{PRL83}, $208(20) \times 10^{12}$ fm$^{-1/2}$,
which corresponds to the black dot of Fig.\ \ref{fits_pwave}(c).
Since the error bars on the phase shifts are very small, a better accuracy on the ANC might even be in reach but this requires to fit
the bell-shaped structure that appears on the experimental background $\Delta_1$ function,
with a maximum reached at 3.7 MeV.
In the following, we assume this structure is physical.
However, it would probably be wise to revisit experimental data,
e.g.\ using a Padé-expanded $\Delta_1$ as a tool for a multienergy phase-shift analysis,
to confirm that this structure is not due to an underestimate of the phase-shift error bars.

Figure \ref{fits_pwave}(c) presents several fits of the data, obtained by a two-step procedure.
First we solve a linear-algebra system providing the coefficients of an $[N_\mathrm{fit}, M_\mathrm{fit}]$ Padé expansion of $\Delta_1$ (the resonance and bound-state energies are allowed to slightly vary), without taking error bars into account.
The total number of free parameters is $N_\mathrm{fit}+M_\mathrm{fit}+1$; they can be formulated as $N_\mathrm{fit}=N_0+N_B+N$ zeros, $M_\mathrm{fit}=N_\infty+M$ poles (possibly complex) and the scattering length $a_l$.
The condition \eqref{CoM} is relaxed as we fit data on a small energy interval.
Second, we start from the obtained parameters to perform a least-square minimization, applied on the randomized data of Refs.\ \cite{PRL88,PRC79}.
An estimate of the ANC error bar $\sigma_\mathrm{ANC}$ can then be obtained from Eq.\ \eqref{ANC_a} and reads,
\begin{equation}
 \frac{\sigma_\mathrm{ANC}}{\mathrm{ANC}} \approx \frac{1}{2} \frac{\sigma_{a_l}}{|a_l|},
 \label{ANCerr_a}
\end{equation}
where $\sigma_{a_l}$ is the uncertainty on the scattering length.
As expected from Fig.\ \ref{fits_pwave}(c), each of these fits provides a very accurate ANC, which shows the power of our method.
For instance, a [2/0] fit [only shown on Figs.\ \ref{fits_pwave}(a) and (b)] on the reduced energy interval [1.955-2.48] MeV
(to get a $\chi^2$ per point smaller than 1) leads to an ANC $=215(1)\times 10^{12}$ fm$^{-1/2}$,
which corresponds to the low-energy plateau of the background $\Delta_1$ function.
However, a [2/1] fit on [1.955-3.88] MeV, which has a stronger slope, leads to the incompatible value $226(1) \times 10^{12}$ fm$^{-1/2}$.
The [2/2], [5/0] and [4/2] fits shown on Fig.\ \ref{fits_pwave} bring $258(2)$, $194(8)$, $223(4) \times 10^{12}$ fm$^{-1/2}$ respectively.
Most fits with other orders lead to unphysical or merging poles and zeros.

This illustrates a difficulty of our method: since it requires an extrapolation on a rather large energy interval, different orders can lead to incompatible ANC values.
In the present case, we use a simplicity argument to combine the ANC estimates of the [2/0], [2/1] and [4/2] fits,
and get the result
 $220.5(6.5) \times 10^{12}$ fm$^{-1/2}$.
Let us finally remark that a Padé approximant of the $\Delta_1$ function allows a direct computation of the scattering-matrix pole in the complex plane, combining Eqs.\ \eqref{Smat} and \eqref{Dlinv}.
All the above approximants lead to a resonance energy $2.3657(4)$ MeV and width $351(2)$ keV in excellent agreement with the similar method developed in Ref.\ \cite{irgaziev15}.

\begin{figure}[]
\scalebox{0.5}{\includegraphics{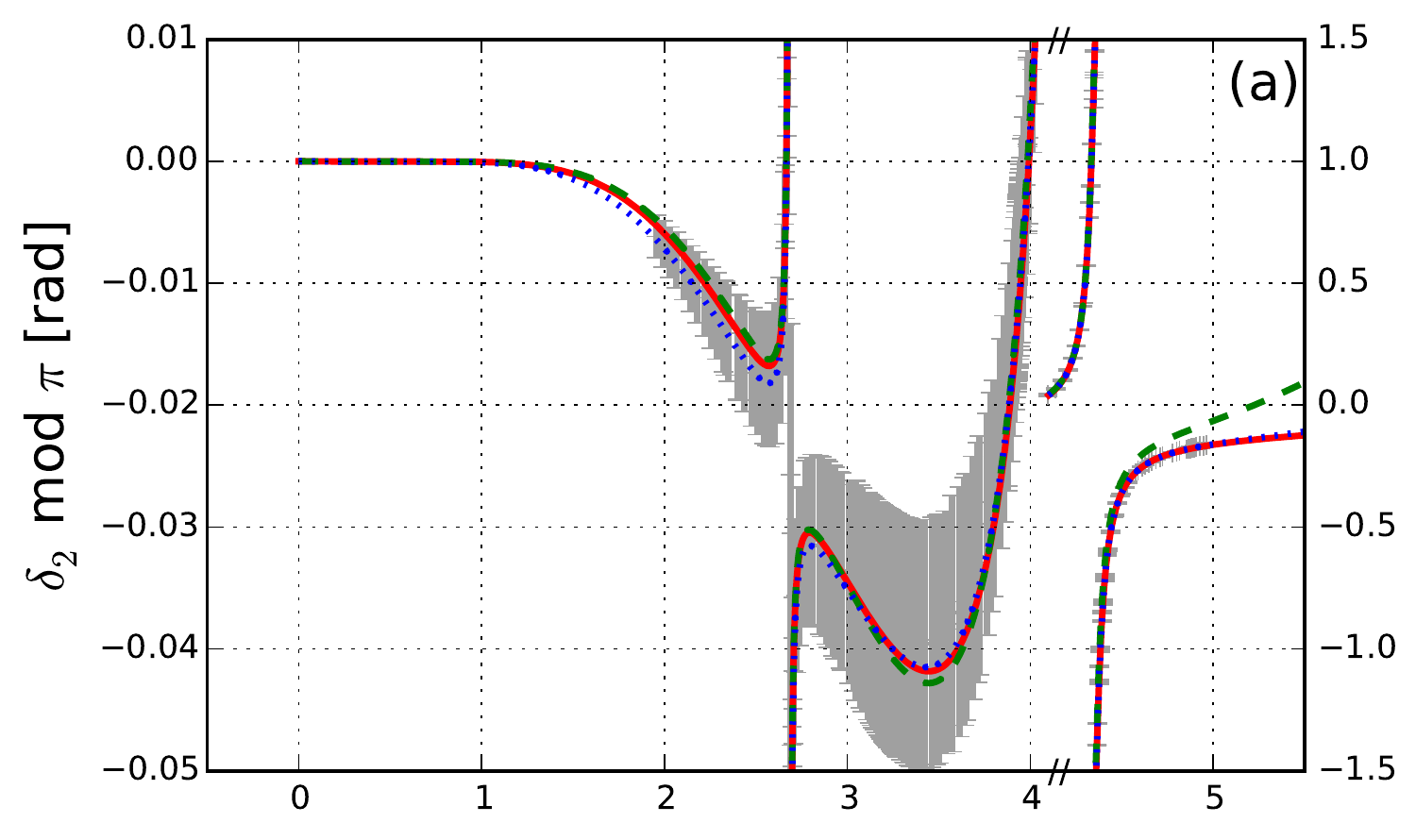}} \\
\scalebox{0.5}{\includegraphics{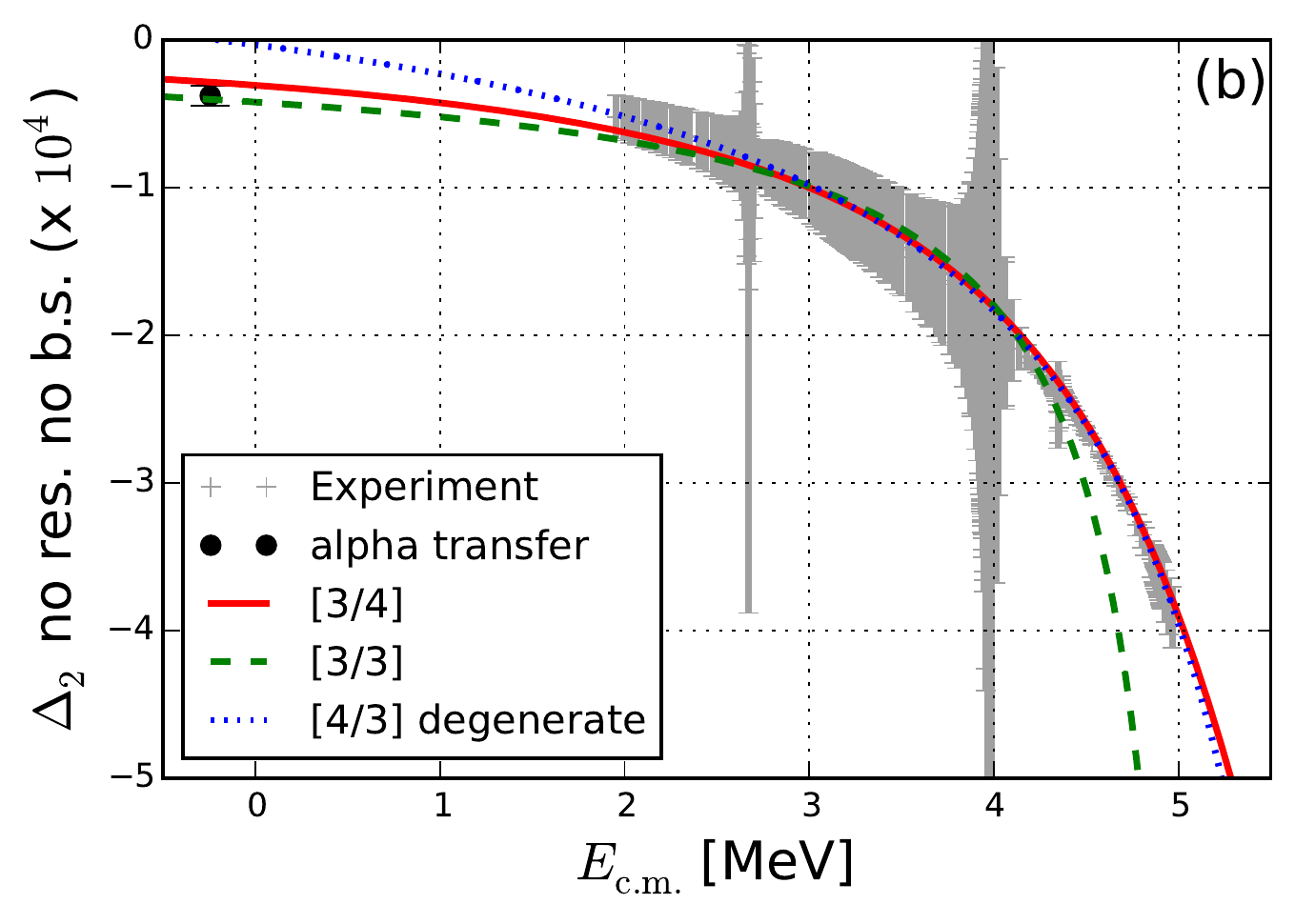}}
\caption{Same as Fig.\ \ref{fits_pwave}(a) and (c) for the $d$ wave, with two resonances ($N_0=N_\infty=2$) and one subthreshold bound-state ($N_B=1$) removed.
The $y$ axis changes at 4.1 MeV in (a).
}
\label{fits_dwave}
\end{figure}

Let us now turn to the $2^+$ wave, for which we follow the same steps.
For this wave, there is a bound state at $E_B=-244.85$ keV and from the experimental phase shifts $\delta_2$ we find $E_{0,j}=\{2.683,\,4.357\}$ MeV and $E_{\infty,j}=\{2.667,\,3.981\}$ MeV, which correspond to the two well-known resonances visible on the experimental phase shifts $\delta_2$ shown on Fig.\ \ref{fits_dwave}(a).
Building the corresponding $\Delta_2$ function and removing these 3 zeros and 2 poles leads to the function plotted in Fig.\ \ref{fits_dwave}(b).
Comparing this figure with the corresponding one for the $p$-wave [Fig.\ \ref{fits_pwave}(c)] shows that the situation is much less favorable here. First, the background $\Delta_2$ function is strongly energy dependent on the experimental energy range. Actually, the best fits obtained for this function favor an additional zero around the threshold energy, which would correspond to an additional bound or resonant $2^+$ state in this region.
This is illustrated for instance by the [4/3] fit of Fig.\ \ref{fits_dwave}, which displays an excellent $\chi^2$ per point of 0.49 on the whole energy range, with only three parameters for the background: one zero close to (or degenerate with) the subthreshold bound state, one pole at about 6.3 MeV and the scattering length.
Though the existence of a degenerate subthreshold state is highly unlikely, testing this hypothesis might be interesting, both from the experimental and theoretical points of view.

Second, despite their good accuracy, the phase shifts of Ref.\ \cite{PRC79} lead to rather large error bars on the background $\Delta_2$ function, as seen on Fig.\ \ref{fits_dwave}(b).
Hence, the extrapolation of this function towards low energies is inaccurate and the prediction for the ANC is not expected to improve on the one of Ref.\ \cite{PRL83}, $\text{ANC}=114(10)\times 10^3$ fm$^{-1/2}$, which corresponds to the black dot of Fig.\ \ref{fits_dwave}(b). Indeed, rather different values for the ANC can be obtained with different fit orders. For instance, the [3/3] and [3/4] fits presented in Fig.\ \ref{fits_dwave}, which present no degenerate bound state because $N=0$, lead to ANCs of $110(11)\times 10^3$ fm$^{-1/2}$ and $131(15)\times 10^3$ fm$^{-1/2}$ respectively. The [3/3] fit has a $\chi^2$ per point of 0.54 but is limited to energies lower than 4.3 MeV and fails to reproduce higher-energy data. Moreover, it reaches the upper end of the phase-shift error bars at low energies; choosing a larger energy interval brings the fit outside these error bars and hence leads to an even smaller value for the ANC. Similar results are obtained for the [4/3] fit forcing the additional zero to stay at large positive energies. The [3/4] fit in contrast is able to fit the data on the whole interval, again with 3 parameters for the background. The corresponding value of the ANC, with its rather large error bar, is compatible both with the values of Refs.\ \cite{PRC79} and \cite{sparenberg04}, hence making any definite conclusion hazardous. Fits with $N=0$ and larger $M$ lead to larger values for the ANC but with even larger error bars.
Let us mention a final difficulty of our method in a case like this, with rather large error bars:
by construction, we only explore one small region of the parameter space at a time.
Hence, only local minima are considered and they are determined by the initial values chosen for the parameters.
These initial values are deduced from the (smooth) R-matrix phase shifts.
We have checked that a more sophisticated method, based on a direct calculation of the Padé-approximant coefficients from randomized data,
leads to essentially the same results for the $d$-wave fits just presented \cite{ramirez14}.

To sum up, our new perspective to analyze phase shifts shows many advantages: (i) it describes $\delta_l$ in a wide energy range, (ii) it is compatible with the S-matrix properties, (iii) it classifies the parameters in two sets, one fixed $\{E_{0},\,E_{\infty},\,E_{B}\}$ with a direct physical meaning (gross phase-shift structure, including resonances and background) and another free $\{p_j,\,q_j\}$ which describes collectively the phase-shift details, (iv) it does not require a channel radius, (v) it does not require a potential
and (vi) it can estimate ANCs directly from elastic phase shifts.
For the $^{12}$C+$\alpha$ phase shifts, the method leads to much better and simpler constraints on the subthreshold-bound-state ANCs than the R-matrix, modified K-matrix or traditional effective-range function. For the $d$-wave, an accuracy similar to the best ones available today is reached, whereas for the $p$ wave a structure hidden up to now is revealed from the data, which would deserve further study, and at least a factor 3 is gained on the accuracy.
In the future, we plan to apply our method to other systems and to extend it to coupled channels and capture reactions.

This text presents research results of the IAP program P7/12 initiated by the Belgian-state Federal Services for Scientific, Technical, and Cultural Affairs, which supported O.L.R.S.\ during his PhD in Brussels. We thank D.\ Baye, C.\ Brune, P.\ Descouvemont, R.\ Johnson, R.\ Raabe and N.\ Timofeyuk for useful discussions at various stages of this work, as well as the Python-language community for many useful programs.


\begin{thebibliography}{30}
\bibitem{Bertulani} C. A. Bertulani and P. Danielewicz, \textit{Introduction to nuclear reactions} (IoP, Bristol and Philadelphia, 2004).
\bibitem{Joachain} C. J. Joachain, \textit{Quantum collision theory} 3rd ed. (North-Holland, Amsterdam, 1999).
\bibitem{buchmann96} L. Buchmann \emph{et al.}, Phys. Rev. C \textbf{54}, 393 (1996).
\bibitem{azuma94} R. E. Azuma \emph{et al.}, Phys. Rev. C \textbf{50}, 1194 (1994).
\bibitem{PRL83} C. Brune \emph{et al.}, Phys. Rev. Lett. \textbf{83}, 4025 (1999).
\bibitem{RPP71} P. Descouvemont and D. Baye, Rep. Prog. Phys. \textbf{71}, 036301 (2010).
\bibitem{PRL88} P. Tischhauser \emph{et al.}, Phys. Rev. Lett. \textbf{88}, 072501 (2002).
\bibitem{PRC79} P. Tischhauser \emph{et al.}, Phys. Rev. C \textbf{79}, 055803 (2009).
\bibitem{sparenberg04} J.-M. Sparenberg, Phys. Rev. C \textbf{69}, 034601  (2004).
\bibitem{NPB60} J. Hamilton, I. Overbö and B. Tromborg, Nucl. Phys. \textbf{B60}, 443 (1973).
\bibitem{PRC29} Z. R. Iwinski, L. Rosenberg and L. Spruch, Phys. Rev. C \textbf{29}, 349 (1984).
\bibitem{PRC81} J.-M. Sparenberg, P. Capel and D. Baye, Phys. Rev. C \textbf{81}, 011601(R) (2010).
\bibitem{JPCS312} J.-M. Sparenberg, P. Capel and D. Baye, J. Phys.: Conf. Ser. \textbf{312}, 082040 (2011).
\bibitem{PRC39} C. R. Chen, G. L. Payne, J. L. Friar and B. F. Gibson, Phys. Rev. C \textbf{39}, 1261 (1989).
\bibitem{blokhintsev93} L. D. Blokhintsev \emph{et al.}, Phys. Rev. C \textbf{48}, 2390 (1993).
\bibitem{pupasov11} A. Pupasov, B. F. Samsonov, J.-M. Sparenberg and D. Baye, Phys. Rev. Lett. \textbf{106}, 152301 (2011).
\bibitem{orlov16} Yu. V. Orlov, B. F. Irgaziev and L. I. Nikitina, Phys. Rev. C \textbf{93}, 014612 (2016).
\bibitem{NPA271} J. Humblet, P. Dyer and B. A. Zimmerman, Nucl. Phys. \textbf{A271}, 210 (1976).
\bibitem{humblet90} J. Humblet, Phys. Rev. C \textbf{42}, 1582 (1990).
\bibitem{mukhamedzhanov99} A. M. Mukhamedzhanov and R. E. Tribble, Phys. Rev. C \textbf{59}, 3418 (1999).
\bibitem{brune96} C. Brune, Nucl. Phys. A \textbf{596}, 122 (1996).
\bibitem{humblet98} J. Humblet, A. Csótó and K. Langanke, Nucl. Phys. A \textbf{638}, 714 (1998). 
\bibitem{burke} P. G. Burke, \textit{R-Matrix Theory of Atomic Collisions} (Springer, Berlin, 2011).
\bibitem{PRC88} O. L.  Ramírez Suárez and J-M. Sparenberg, Phys. Rev. C \textbf{88}, 014601 (2013).
\bibitem{PRC61} D. Baye and E. Brainis, Phys. Rev. C \textbf{61}, 025801 (2000).
\bibitem{PRC63} D. Baye, M. Hesse and R. Kamouni, Phys. Rev. C \textbf{63}, 014605 (2000).
\bibitem{PRA26} H. van Haeringen and L. Kok, Phys. Rev. A \textbf{26}, 1218–1225 (1982).
\bibitem{ramirez14} O. L. Ramírez Suárez, PhD thesis, Université libre de Bruxelles (2014).
\bibitem{irgaziev15} B. F. Irgaziev and Yu. V. Orlov, Phys. Rev. C \textbf{91}, 024002 (2015).

\end{thebibliography}
\end{document}